\RequirePackage{fix-cm}
\documentclass[twocolumn]{svjour3}          % twocolumn
\smartqed  % flush right qed marks, e.g. at end of proof
\usepackage{graphicx}
\usepackage[authoryear]{natbib}
\usepackage{multirow}
\def\makeheadbox{{%
		\hbox to0pt{\vbox{\baselineskip=10dd\hrule\hbox
				to\hsize{\vrule\kern3pt\vbox{\kern3pt
						\hbox{\bfseries Journal of Intelligent Manufacturing}
						\hbox{This is a pre-print version of the submitted article.}
						\kern3pt}\hfil\kern3pt\vrule}\hrule}%
			\hss}}}

\begin{document}

\title{Fully Convolutional Networks for Chip-wise Defect Detection Employing Photoluminescence Images 
\thanks{The authors would like to thank OSRAM Opto Semiconductors and especially Dr. Hans Lindberg for data provision, ongoing support and a great collaboration.}
}
\subtitle{Efficient Quality Control in LED Manufacturing}

\titlerunning{FCNs for Chip-wise Defect Detection}        % if too long for running head

\author{Maike Lorena Stern        \and
Martin Schellenberger
}

\institute{\at
Fraunhofer Institute for Integrated Systems and Device Technology IISB \\
Schottkystrasse 10 \\
91058 Erlangen, Germany \\
\\
\emph{Present address of first Author:}   \\
OSRAM Opto Semiconductors \\
Leibnizstrasse 4 \\
93055 Regensburg, Germany \\
E-Mail: Maike.Stern@osram-os.com
}

\date{Received: date / Accepted: date}
% The correct dates will be entered by the editor

\maketitle

\begin{abstract}
\begin{sloppypar}
Efficient quality control is inevitable in the manufacturing of light-emitting diodes (LEDs). Because defective LED chips may be traced back to different causes, a time and cost-intensive electrical and optical contact measurement is employed. Fast photoluminescence measurements, on the other hand, are commonly used to detect wafer separation damages but also hold the potential to enable an efficient detection of all kinds of defective LED chips. On a photoluminescence image, every pixel corresponds to an LED chip's brightness after photoexcitation, revealing performance information. But due to unevenly distributed brightness values and varying defect patterns,  photoluminescence images are not yet employed for a comprehensive defect detection. In this work, we show that fully convolutional networks can be used for chip-wise defect detection, trained on a small data-set of photoluminescence images. Pixel-wise labels allow us to classify each and every chip as defective or not. Being measurement-based, labels are easy to procure and our experiments show that existing discrepancies between training images and labels do not hinder network training. Using weighted loss calculation, we were able to equalize our highly unbalanced class categories. Due to the consistent use of skip connections and residual shortcuts, our network is able to predict a variety of structures, from extensive defect clusters up to single defective LED chips. 
\end{sloppypar}

% \PACS{PACS code1 \and PACS code2 \and more}
% \subclass{MSC code1 \and MSC code2 \and more}
\end{abstract}

\section{Introduction}
\label{intro}

\begin{sloppypar}
The manufacturing of LEDs is a complex semiconductor process, interspersed with various measurements. These measurements serve as a means to monitor production steps as well as to determine wafer properties, subsequently bin LED chips and separate out flawed ones, respectively. However, because measurements are not value adding, LED manufacturers have an interest in reducing their deployment as far as possible. 

\begin{figure*}[h]
\centering
\includegraphics[width=.9\linewidth]{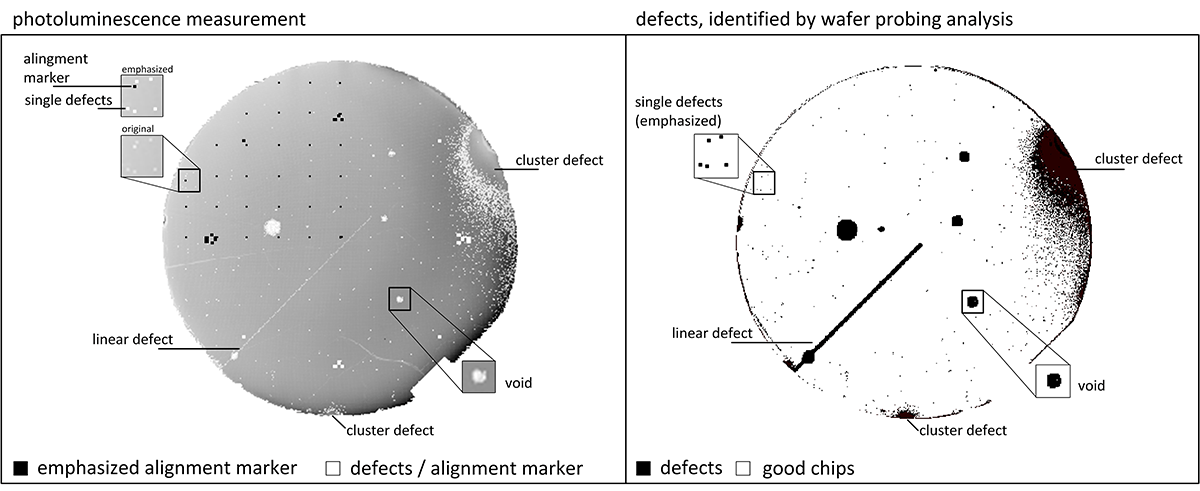}
\caption{Photoluminescence image of a wafer with 133,717 LED chips (left) and the corresponding probing-based defect map (right). Defects depicted on the defect map are also visible on the photoluminescence image, independently of the determined defect cause. We classify defect types by their appearance on the photoluminescence image, namely single defects, linear defects, voids and defect clusters. Notice that the photoluminescence image also displays alignment markers (for improved recognizability, as they resemble defects, emphasized in black on a fourth of the wafer). Also, voids tend to appear smaller on photoluminescence measurements compared to the probing-based labels because ultrasonic measurements, which are included in the prober defects, determine the actual air pocket within the wafer structure while photoluminescence measurements can only depict the corresponding flawed optical surface.}
\label{fig:pl-probing}
\end{figure*} 

Wafer probing, for example, determines the electrical and optical chip properties by contacting each and every chip on the wafer with a prober needle and performing a series of tests; it is hence a precise but time-consuming measurement. Moreover, decreasing chip sizes give rise to even more chips per wafer, increasing measurement time further. Less precise but faster photoluminescence measurements, on the other hand, determine only the chip's brightness: here, the chip's optical surface is being radiated by photo-excitation and thus the measured brightness is dissimilar from the one emitted by the entire LED, induced by electrical excitation. Still, comparing photoluminescence measurements with probing-based defect maps reveals that conspicuous LED chips can be identified on photoluminescence images as well, as shown in figure~\ref{fig:pl-probing}. Notice that all electrical and optical defect causes generated by wafer probing have been subsumed into one defect class. Henceforth, we will classify defect types by their appearance on the photoluminescence image (independently of their defect cause) as they are differently hard to recognize for the algorithm. Defect types range from single defective LED chips -- usually visible as salient pixel -- to assemblages of defects, namely linear defects, voids and defect clusters. 
\end{sloppypar}

\begin{figure*}
\centering
\includegraphics[width=1.\linewidth]{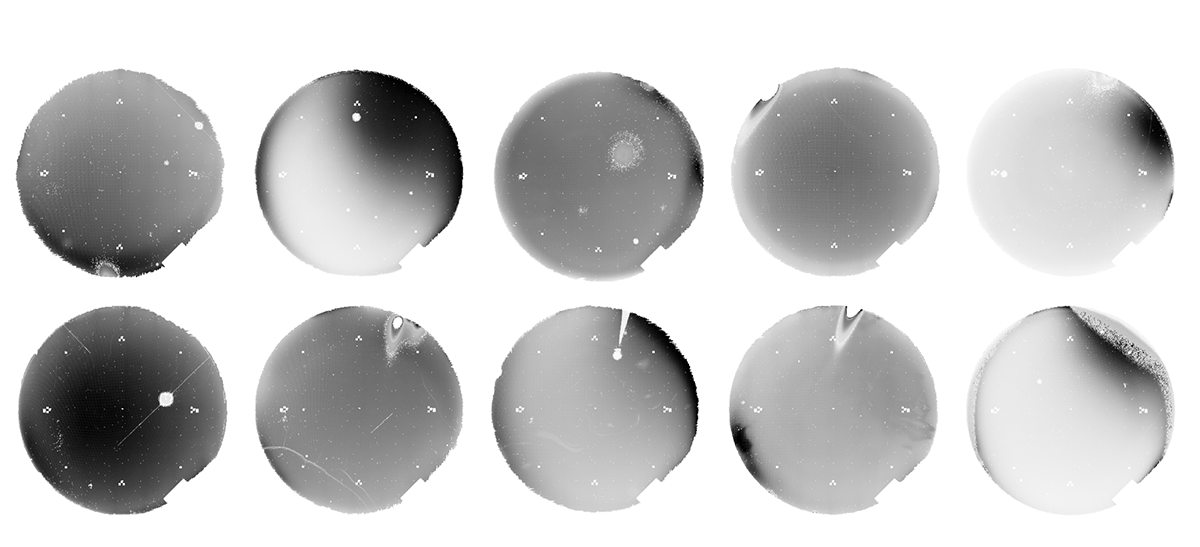}
\caption{Overview over the variety of photoluminescence measurement results with different brightness values as well as defect cluster shapes and sizes. Due to the measurement setup brightness values can differ significantly within a wafer as well as from wafer to wafer. Additionally, defect clusters assume different sizes and shapes along with an interchange of sharp and smooth brightness gradients.}
\label{fig:multiplewafers}
\end{figure*}

While data analysis methods -- such as thresholding the brightness of adjacent LED chips -- can often detect single defects, other defect patterns may elude recognition due to the non-uniformity of brightness values. Figure~\ref{fig:multiplewafers} gives an overview of possible measurement results: on the one hand, photoluminescence measurements can yield unevenly distributed results that are independent of the LED chip's genuine brightness and can exceed brightness differences within defect areas (see especially fig. \ref{fig:multiplewafers}, top right). On the other hand, defect clusters assume diverse sizes and shapes and additionally can display an interchange of sharp and smooth brightness gradients within the defect area. Furthermore, defect clusters occur rarely in LED manufacturing, making yet unknown, extraordinary cluster shapes probable and hence creating a further barrier to capturing all possible cluster features in one data-set. It is therefore not reasonable to hand-craft an algorithm that employs heuristic features, which is why we developed a self-learning defect classification algorithm.

\begin{sloppypar}
In recent years, Convolutional Neural Networks (CNNs) are driving major advances in a variety of computer vision tasks, such as image classification \citep{Krizhevsky,Simonyan,Szegedy,ResNets}, object detection \citep{Ren,Yolo} and semantic image segmentation \citep{Kraehenbuehl,Long,Segnet,Ronneberger,Yu,Lin,Chen,Zhao}. Here, semantic segmentation refers to the task of allocating each pixel to a fixed set of class categories, which in our case corresponds with classifying every single LED chip. The main advantage of CNNs, in comparison to many other data analysis techniques, is their ability to learn shared parameters in form of convolution filters. These filters are slid across the input image and are hence capable of detecting defect structures that may occur on different image positions. Also, CNNs obviate the need for hand-crafted features as they learn the suitable filter parameters for the task through backpropagation. To output a set of class categories, CNNs heavily reduce the input image with so called pooling layers and thereby lose spatial information. Because the pooling operation has additional advantages, such as keeping the computational requirements reasonable, skipping them is not a means to implement pixelwise classification. 
\end{sloppypar}

Fully convolutional networks (FCNs) \citep{Long,Shelhamer}, on the other hand, offer a modified CNN architecture that recovers the spatial resolution by introducing in-network upsampling layers. Now, the network is composed of a downsampling and a succeeding upsampling part. In order to enhance the coarse information resulting from upscaled low-resolution images, additional skip connections are introduced. Skip connections fuse outputs of shallower layers with deeper layers and thereby refine the spatial precision of the output. Among all methods for pixelwise predictions, FCNs were the first to enable end-to-end training as well as the deployment of transfer learning and have since been refined by several working groups \citep{Tiramisu,Li,Abdulnabi,deepUnet,Ronneberger}. 

\begin{sloppypar}
Within an industrial environment, large, labeled data-sets are rare. Here, transfer learning provides a means to accelerate learning by re-utilizing network weights, pre-trained on standard data-sets containing hundreds of thousands of pictures \citep{transferLearning,Imagenet}. Due to the basic elements all images share, such as edges, changes in brightness, etc., representations learned from a distinctly larger data-set can help to generalize from a small data-set. However, we noticed that due to the simplicity of our data's features, transfer learning has only a limited impact on our network's learning speed and accuracy, as shown in section \ref{sec:experiments}.
\end{sloppypar}

To the best of our knowledge, this is the first work that presents a fully convolutional network for the fast, chip-wise defect detection of LED chips on a wafer, based on photoluminescence images. Pixel-wise labels have been created using wafer probing results, being the most accurate chip-wise classification available. The developed network architecture was especially designed with regard to the specific image composition of photoluminescence images as well as the need for a pixel-accurate prediction resolution and lays the foundation for the employment of fully convolutional networks in an industrial production. We report the performance of our method quantitatively and qualitatively on a data-set of 145 wafers.

\section{Related Work}
\label{sec:1}

Due to the specific application of this task we are not aware of any reported previous work with regard to pixel-wise classification of defects in photoluminescence measurements. The application of fully convolutional networks can be found in various other areas, such as medical image analysis and scene understanding.  
As an instance, several groups have reported on automatic segmentation methods for biomedical image analysis, in order to render time-consuming, manual segmentations by trained experts unnecessary \citep{Ronneberger,Tai,Tran,Sharma}. The understanding of street scenes with regard to pedestrians and environment or road detection is a crucial aspect of autonomous driving and driver assistance systems \citep{Pohlen,Uhrig}. \cite{Demant} outline several methods to evaluate the quality of multi-crystalline silicon solar cells using photoluminescence measurements, while \cite{Lin2018} report on a convolutional neural network that distinguishes two different defect types in single LED chip images, where the defect regions are localised with bounding boxes. However, both works do not implement pixel-wise semantic segmentation or fully convolutional networks.

\section{Methods}
\label{sec:methods}

\begin{figure}
\centering
\includegraphics[width=1. \linewidth]{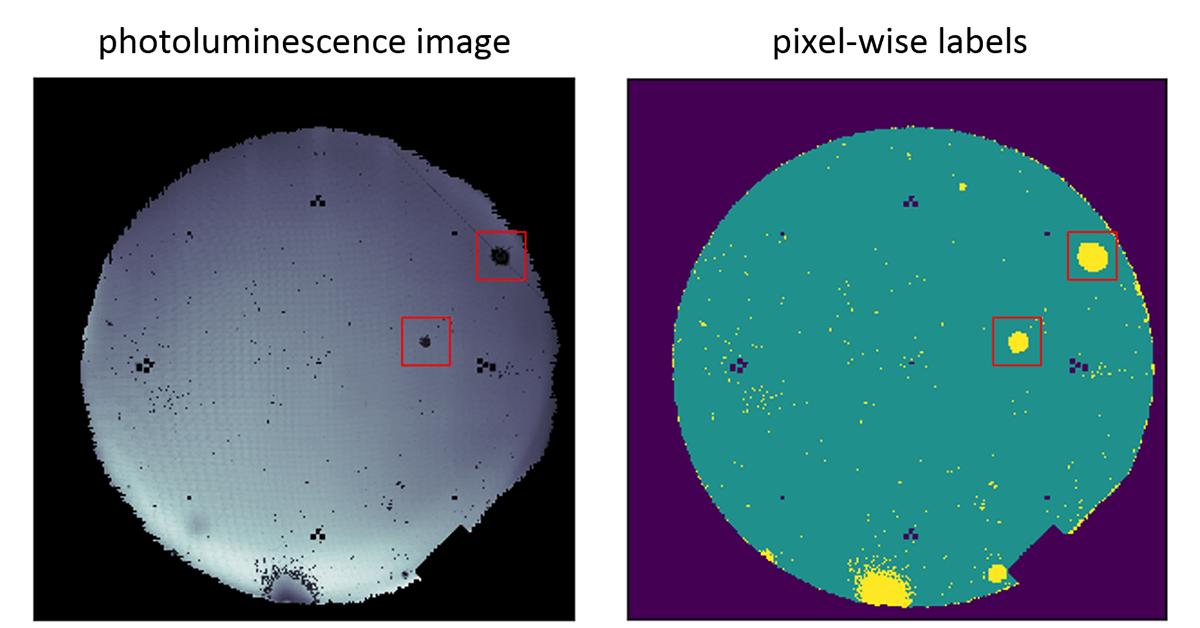}
\caption{Photoluminescence image and corresponding labels, where every pixel of the photoluminescence image is assigned a class category. Yellow depicts defects, turquoise stands for in-spec chips and dark blue for background pixels as well as alignment markers. The red squares illustrate the different depiction of voids in photoluminescence images and labels, respectively. Best viewed in color.}
\label{fig:labels}
\end{figure}

\paragraph{Photoluminescence Image and Label Acquisition}
\begin{sloppypar}
Data-sets for semantic segmentation depend on the possibility to obtain pixel-wise labels. In our case, the measurement we actually intend to omit provides an accurate classification of every single LED chip, which we can use to create pixel-wise class labels (see next paragraph). The probing defect map covers not only probing results based on electrical and optical measurements; supplementary data evaluations and measurements conducted earlier in the process are included as well. As an instance, ultrasonic measurements are performed prior to wafer probing and determine structural damages below the wafer surface (voids, e.g.) so as to spare those flawed areas while probing. The scope of probing-based defect information can lead to discrepancies between photoluminescence image and label: the aforementioned voids, for example, can be determined by ultrasonic and photoluminescence measurements alike, as voids usually result in dark spots visible on photoluminescence images but with a smaller defect area than given by ultrasonic labels (see figure \ref{fig:labels}, red squares). Because we were interested in the network's performance with regard to unaltered, probing-based labels, we initially refrained from enhancing photoluminescence images with any additional information. As described in section \ref{sec:experiments}, our experiments indicate that image-label discrepancies do not hinder network training but can constrain the validity of performance metrics. Experiments using photoluminescence images with embedded ultrasonic measurement results consequently show increased defect class accuracy.
\end{sloppypar}

\paragraph{Class Categories and Wafer Selection}
As mentioned, wafer probing constitutes a compilation of several tests and evaluations, yielding over 20 defect classes. Because the first failure that occurs during the probing sequence is assigned as defect cause, defect classes do not necessarily correspond with the core reason for failure. Additionally, defect causes might have been assigned before wafer probing, ultrasonic and visual measurement results as an instance. We thus subsumed all defect causes into one class (class 2, figure \ref{fig:labels}: yellow) and created two additional classes, one representing in-spec chips (class 1, turquoise) and the other one representing background as well as alignment markers (class 0, dark blue). The need for a background class emerges from the embedding of the round wafer area into a rectangular image format. For our experiment we used LED wafers consisting of 133,717 LED chips, resulting in a $ 442 \times 440 $ pixel matrix. Usually, only a fractional amount of all LED chips are classified as defective. Among different defect patterns---such as single defects, voids, linear defects and defect clusters---clusters occur the rarest and are the most difficult to recognize. Hence, we focused on wafers with defect clusters, of which we have 54 wafers at our disposal, and extended our data-set with additional 91 wafers displaying prominent defect patterns, resulting in a data-set of 145 wafers. 

\paragraph{Network Architecture}

\begin{figure*}[h]
	\centering
	\includegraphics[width=.9\linewidth]{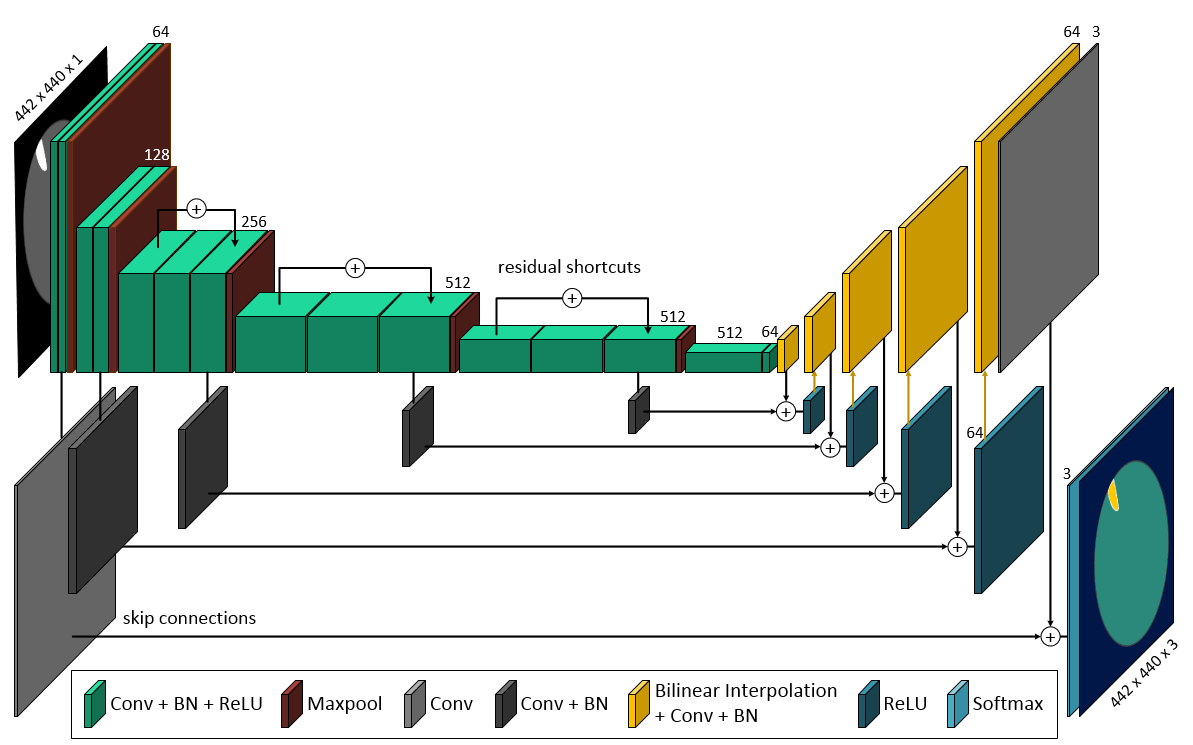}
	\caption{Depiction of our fully convolutional network architecture. The CNN part is adapted from the VGG\,16 network \citep{Simonyan}, but with an abbreviated last CNN layer stack of only 512 and 64 feature maps, respectively. The upsampling part is composed of fixed bilinear interpolation and subsequent convolution with a layer depth of 64. Residual shortcuts ease training and skip connections enable fine-grain predictions. Architecture portrayal adapted from \cite{Tai}.}
	\label{fig:architektur}
\end{figure*}

FCNs consist of three parts: a CNN part that inputs the images and converts them into semantic information, an upsampling part that recovers spatial information as well as skip connections to enhance the resolution. In order to accelerate training by re-utilizing pre-trained weights (transfer learning) it is necessary to adapt the CNN architecture of the weight giving network. Due to its straightforward structure and freely shared parameters, the so-called VGG\,16 network, developed by Karen Simonyan and Andrew Zisserman, is a common choice \citep{Simonyan}. During our experiments, we noticed that in our case the learning advantage of a partly pre-trained network is caught up by a randomly initialized network after only 25 epochs, where we use a normal distribution as described by \cite{HeInit}. Also, both initialization methods achieve similar pixel accuracies with 98.7\,\% (VGG init.) to 98.6\,\% (He init.). With regard to defect class accuracy, however, we observed that employing transfer learning does have a more distinct influence, with 54.7\,\% defect class accuracy (VGG init.) to 51.4\,\% (He init.). We assume that the simple features of class 0 (background) and class 1 (in-spec chips) can be learned from scratch just as fast as it takes the network to re-build VGG\,16 filters, whereas the more complicated defect pattern features of class 2 benefit from pre-trained filters. Our experiments indicate that the learning process gains the most from a transfer of similar filters and a random initialization of the remaining ones, as shown in section \ref{sec:experiments}. 

The CNN architecture of our network follows the VGG\,16 representation, including a continuous, small receptive field of $ 3 \times 3 $ and zero padding. Because the last CNN layer serves as dimension reduction, we employ a $ 1 \times 1 $ filter. In order to ease training, we perform batch normalization (BN) after every convolutional operation. By doing so, the internal covariance shift problem is avoided by normalizing every batch to zero mean and unit variance \citep{Ioffe}. To introduce non-linearity, every BN layer is followed by a rectified linear unit (ReLU) layer as activation function \citep{Nair}. The consecutive operations convolution, batch normalization and rectified linear unit (Conv, BN, ReLU) are shown as green block in figure \ref{fig:architektur}, where the number over the block represents the number of filters and feature maps, respectively. After every layer stack of two or three layers a dimension reduction is performed, using max-pooling (red) with $ 2 \times 2 $ filters and stride 2.  

VGG\,16 is a convolutional neural network and thus consists of 13 fully convolutional and three fully connected layers. As outlined in \cite{Shelhamer}, fully connected layers can be transformed into (fully) convolutional layers, resulting in three additional convolutional layers with 4096, 4096 and 1000 filters, respectively. As shown by \cite{Zeiler} and \cite{Yosinski}, filters in successive layers combine features of previous filters, thus creating more complex patterns with increasing network depth. In order to omit layers that are overly complex for the data's patterns, we abbreviated this last convolutional layer stack (conv6), resulting in two layers with 512 (conv6\_1) and 64 filters (conv6\_2), respectively. Furthermore, we added residual shortcut connections \citep{He} to all threefold stacked layers (conv3, conv4, conv5), so as to ease the approximation of identity mappings and hence allow the network to further adjust the number of filters itself. While shortcuts had no effect on pixel accuracy they did improve mean pixel accuracy (83.7\,\% to 84.5\,\%) as well as defect class accuracy (52.6\,\% to 54.7\,\%).

After downsampling, the image's spatial dimensions must be restored through upsampling. Usually, a transpose convolution operation (deconvolution) is employed in order to reverse the previous convolution. We noticed, however, that due to our small data-set transpose convolution operations give rise to increased overfit, compared to a fixed upsampling operation. We therefore resize the images with bilinear interpolation and a subsequent convolutional layer (yellow block, figure \ref{fig:architektur}) \citep{distill}. As shown in figure \ref{fig:architektur}, the resizing steps follow the downsizing steps of the CNN part, which allows the implementation of skip connections. These shortcuts help recovering fine-grained information by re-using feature maps of shallower, less downsampled layers. In our experiments, we achieved the best results with five skip connections, where each CNN layer stack gets connected with its upsampling counterpart. 

We set the number of feature maps throughout the upsampling path to 64 and reduced the depth of the skip layers accordingly with $ 3 \times 3 $ convolutional filters. Then, after adding skip layer (grey) and resized layer (yellow), we apply a ReLU activation function before extending image dimensions until the original image size of $ 442 \times 440 $ is obtained. In the last hidden layer the number of feature maps is reduced to the number of class categories by applying $ 3 \times 3 $ convolutional filters on both, skip layer as well as resized layer. Then, after summarizing the layers, a softmax activation function calculates the per-pixel likelihood score for all semantic categories. In order to generate a prediction map, each pixel is assigned the class with the highest probability.

\paragraph{Data Augmentation, Training and Validation}
In order to create a balanced training and validation set, we split our data-set of 145 wafers into 106 training images and 39 validation images, where the validation set includes eight wafers with defect clusters and the training set 46 wafers. Due to the small number of wafers with cluster defects we divided them hand-picked to make sure that the training data contains an overview of available defect clusters and the validation set contains at least one exceptional cluster shape. As safeguard, we also performed random 4-fold cross validation (see section \ref{tab:CrossValidation}). To further extend our training data-set, we employed data augmentation to improve network generalization and mitigate overfitting. Since neither the image object nor the image quality are prone to changes we applied $ 90^\circ$, $ 180 ^\circ $ and $ 270 ^\circ$ rotations only, resulting in 424 training images. To prepare the data, we executed several pre-processing steps, including the transformation into a network readable format, image normalization and one-hot-encoding the labels. Additionally, the mean pixel value of the VGG\,16 network was subtracted and the data was shuffled randomly before every epoch. Using a batch size of 1, training was performed on an NVIDIA GeForce GTX 1080 Ti (11~GB) GPU and an Intel Xeon 3.20 GHz CPU. Training the network end-to-end, for 200 epochs, took about four hours. All experiments were performed using the Tensorflow framework \citep{tensorflow}. In order to optimize the network during training, we calculate the deviation between prediction and true label with a cross entropy loss function. Because defects occur rarely---compared to in-spec chips and background pixels---we equalized loss calculation by applying a weight of 2000 on the defect class and a weight of 100 to the other two classes (see section \ref{sec:experiments}). Independent of the initialization method, we fine-tuned all layers by backpropagation through the whole network and in one stage, using adaptive moment estimation (Adam) \citep{Adam} and an initial learning rate of 0.0008, which was decayed exponentially. Network regularization was implemented using a constant weight decay of 0.0005.

\paragraph{Metrics}
\begin{sloppypar}
To evaluate network performance, we report metrics common for semantic segmentation, namely pixel accuracy, mean pixel accuracy and mean region intersection over union (IoU). Mean pixel accuracy determines an improved pixel accuracy by computing the ratio of correct pixels per class and then averaging over the total number of classes. Mean IoU is the standard metric for segmentation and determines the averaged ratio between the intersection and the union of two sets, that is the true and the predicted segmentation. The metrics are calculated as follows:
\end{sloppypar}

\begin{itemize}
\renewcommand{\labelitemi}{$\bullet$}
\item pixel accuracy:  $ {\sum\nolimits_{i} p_{ii}} \, / \, {\sum\nolimits_{i} t_i} $
\item mean pixel accuracy: $ (1/n_c) \, {\sum\nolimits_{i} p_{ii}} \, / \, {t_i} $
\item mean IoU: $ (1/n_c) \, {\sum\nolimits_{i} p_{ii}} \, / \, (t_i + \sum\nolimits_{j} p_{ji}-p_{ii})$
\end{itemize}

\noindent where $ p_{ji} $ is the amount of pixels of class $ j $ predicted to belong to class $ i $, $ t_i = \sum\nolimits_{j} p_{ij} $ is the total number of pixels in class $ i $ and $ n_c $ is the number of classes. In order to take the sparsity of class objects pertaining to the defect class into account, we also calculated the pixel accuracy of the defect class alone.

\section{Experiments and Results}
\label{sec:experiments}

\paragraph{Architecture}

\begin{table*}[h]
\centering
% table caption is above the table
\caption{Overview over the three tested architecture variants. The standard structure (network 1) follows the VGG\,16 setup in its CNN part, while network 2 (Vaughan) is equipped with an abbreviated last CNN layer stack of only 512 and 64 feature maps, respectively. Network 3 (broomstick) completely omits the last CNN layer stack according with the first upsampling stage.}
\label{tab:architectures}       
% For LaTeX tables use
\begin{tabular}{lllll}
\hline\noalign{\smallskip}
layer stack & output size & \multicolumn{3}{c}{architecture}  \\
\noalign{\smallskip}\hline\noalign{\smallskip}
&& 1: standard & 2: Vaughan & 3: broomstick \\
\noalign{\smallskip}\hline\noalign{\smallskip}
conv1\_x  & 442, 440 & 64, 64  & 64, 64  & 64, 64  \\
conv2\_x &  221, 220 &   128, 128  & 128, 128  & 128, 128  \\
conv3\_x & 111, 110 &  256, 256, 256  &    256, 256, 256 &    256, 256, 256 \\
conv4\_x & 56, 55 &   512, 512, 512 & 512, 512, 512 & 512, 512, 512 \\
conv5\_x & 28, 28 &  512, 512, 512 & 512, 512, 512 & 512, 512, 64 \\
conv6\_x & 14, 14 & 4096, 4096, 64 & 512, 64 & - \\
trans1 + skip5\_3 & 28, 28 & 64 &  64 & - \\
trans2 + skip4\_3 & 56, 55 &  64 & 64 & 64 \\
trans3 + skip3\_3 & 111, 110 &   64 & 64 & 64  \\
trans4 + skip2\_2 & 221, 220 &  64 & 64 & 64 \\
trans5 + skip1\_2 & 442, 440 &   64 & 64 & 64  \\
\noalign{\smallskip}\hline
\noalign{\smallskip}
\end{tabular}
\end{table*}

\begin{sloppypar}
To choose the suitable network design for our data, we evaluated performance metrics of different architectures and hyperparameters. At first, we compared three architecture variants, where the basic structure corresponds with the architecture described in section \ref{sec:methods}. Proceeding from VGG\,16's layer structure (architecture 1, standard), we tested two abbreviated versions: architecture 2 (Vaughan) is composed of the same number of layer stacks as network 1, but counts only 512 and 64 filters in the sixth layer stack, respectively. Architecture 3 (broomstick) omits the last CNN layer stack as well as the first transpose layer completely (table \ref{tab:architectures}). Comparing the training results in table \ref{tab:results_1} shows that the abbreviated network architectures achieve comparable prediction accuracy. This indicates that a full VGG\,16 architecture actually overfits the data but due to residual shortcuts and skip connections the network is able to bypass redundant layers. This assumption is supported by the observation that network 1 (standard) achieves only 90.7\,\% pixel accuracy without skip connections and residual shortcuts whereas network 2 (Vaughan) yields 96.3\,\% accuracy.  
\end{sloppypar}

\begin{table}[h]
\centering
% table caption is above the table
\caption{Comparison of performance metrics of the three network architectures and with three different initialization methods. Pixel accuracy (PA), mean intersection over union (mIoU), mean pixel accuracy (MPA) and defect class accuracy (DCA) are determined on our validation data-set. The results indicate that an abbreviated network architecture combined with a parameter transfer of the first four VGG\,16 layers yields the highest prediction accuracy. }
\label{tab:results_1}       % Give a unique label
% For LaTeX tables use
\begin{tabular}{lllll}
\hline\noalign{\smallskip}
network & PA  & mIoU & MPA & DCA \\
\noalign{\smallskip}\hline\noalign{\smallskip}
% VGG-Gewichte bis 10
\multicolumn{5}{c}{Init. with VGG\,16 weights layer 1 to 10} \\
\noalign{\smallskip}\hline\noalign{\smallskip}
1 standard & 98.6  & 79.7 & 84.1 & 53.7 \\
2 Vaughan & \textbf{98.7}  & 79.8 & 84.1  & 53.4  \\
3 broomstick & \textbf{98.7} & 79.4 & 83.9 & 52.6 \\
\noalign{\smallskip}\hline\noalign{\smallskip}
% VGG-Gewichte bis 2
\multicolumn{5}{c}{Init. with VGG\,16 weights layer 1 to 4} \\
\noalign{\smallskip}\hline\noalign{\smallskip}
1 standard & 98.7  & 80.0  & 84.1  &  53.9  \\
2 Vaughan   & \textbf{98.7}  & \textbf{80.2} & \textbf{84.5} & \textbf{54.7} \\
3 broomstick  & \textbf{98.7}  & 80.1 & 84.2 & 53.7   \\
\noalign{\smallskip}\hline\noalign{\smallskip}
% ohne VGG-Gewichte 
\multicolumn{5}{c}{He initialization} \\
\noalign{\smallskip}\hline\noalign{\smallskip}
1 standard & 98.6 & 79.2& 83.4 & 51.7 \\
2 Vaughan  & 98.6  & 79.1 & 84.3 & 51.4  \\
3 broomstick & 98.6 & 79.0 & 83.1 & 50.5 \\
\noalign{\smallskip}\hline
\end{tabular}
\end{table}

\begin{figure*}[h]
\centering
\includegraphics[width=1.\linewidth]{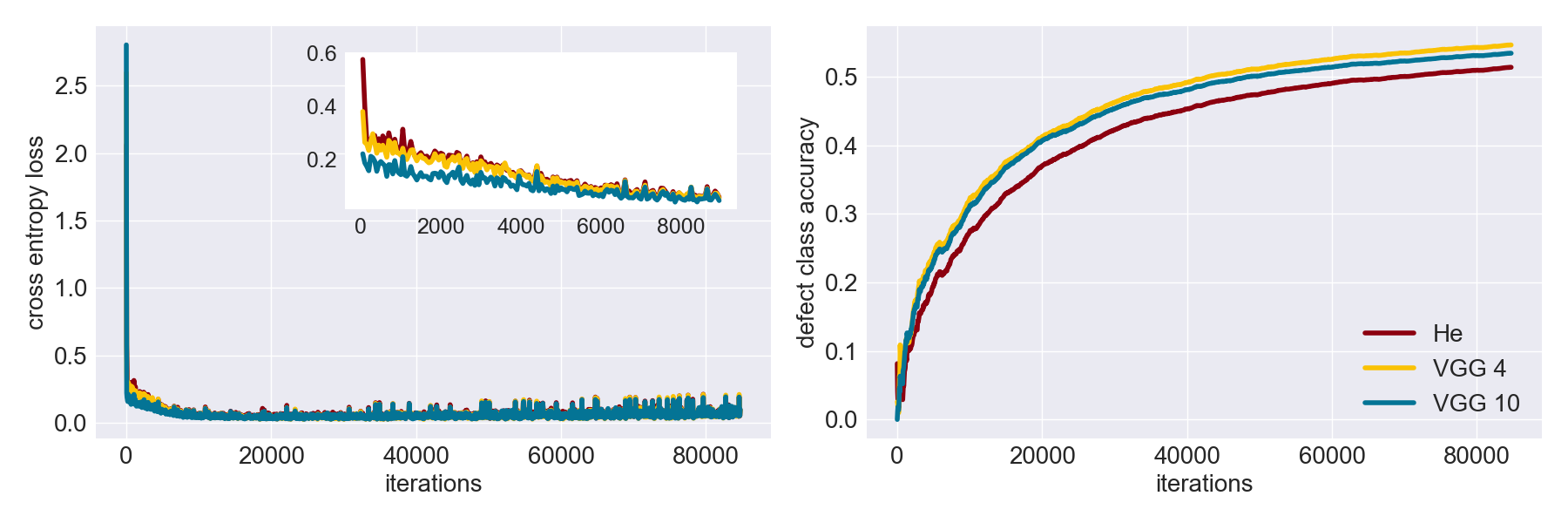}
\caption{Loss and defect class accuracy of the validation data-set resulting from three different initialization methods. Initializing the first ten network layers with VGG\,16 parameters accelerates training (VGG\,10) but full He initialization and an initialization where only the first four layers are provided with VGG\,16 weights (VGG\,4) catch up around step 10,000 (epoch 25). The highest defect class accuracy is achieved with VGG\,4 initialization, followed by the VGG\,10 initialization, indicating that pre-trained filters are more easily adjusted to a new task than randomly initialized filters.}
\label{fig:Init_Vergleich}
\end{figure*}

\paragraph{Initialization}
\begin{sloppypar}
Table \ref{tab:results_1} also lists different initialization approaches: next to randomly initializing the complete network from scratch with He initialization \citep{HeInit}, we transfered VGG\,16 parameters into the first ten (VGG\,10) and the first four (VGG\,4) CNN layers, respectively. Figure \ref{fig:Init_Vergleich} illustrates loss as well as defect class accuracy of network 2 (Vaughan). While the VGG\,10 version initially learns faster, beginning with iteration 10,000 (about 25 epochs) all networks yield comparable loss as well as accuracy. Indeed a difference emerges in defect class accuracy, where the VGG\,4 initialization achieves the highest value (54.7\,\%), followed by VGG\,10 (53.4\,\%) and He initialization (51.4\,\%). This indicates that network learning benefits from the transfer of suitable filters whereas inapt filters slightly dampen learning in comparison to random initialization.  
\end{sloppypar}

% visualise_initialisation_paper2.py

\begin{figure*}
\centering
\includegraphics[width=1.\linewidth]{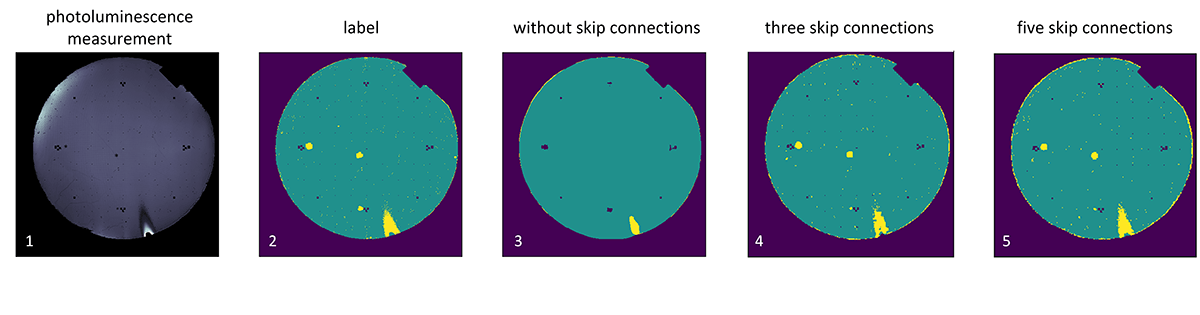}
\caption{Comparison of the network's prediction resolution without skip connections (image 3), with three skip connections (image 4) and with skip connections linking all CNN layer stacks to their upsampling part (image 5). It becomes apparent that skip connections enable a more fine-grain prediction and hence an increased defect class accuracy.}
\label{fig:skipvergleich}
\end{figure*}

\paragraph{Skip Connections}
The network's capability of predicting small defect structures, such as single defective LED chips, depends on the spatial resolution, which is lost due to repeated downsampling in the CNN part of the network. To retrieve the fine-grain information of shallower layers we implement skip connections \citep{Long}. Figure \ref{fig:skipvergleich} illustrates the prediction resolution without (image 3), with three (image 4) and with five skip connections (image 5): while a network architecture without skip connections does detect extensive defect areas, smaller defects and single alignment markers are not identified. Also, the cluster area is represented poorly. Introducing three skip connections to the inner core of the network (conv 4/5/6 to trans 1/2/3) significantly improves prediction resolution and accuracy (image 4). Now, the network depicts defects of all sizes as well as alignment markers in their actual shape. The prediction of the defect cluster area follows the labels more closely but is frayed and holey. Linking all CNN stages to their upsampling counterpart (image 5) improves the defect cluster representation. These pictorial results reflect performance metrics, as a skip-free network achieves a defect class accuracy of 15.3\,\%, while a network with three skip connections reaches 47.3\,\% and a network with five skip connections rises to 54.7\,\% defect class accuracy. 

\begin{figure*}
\centering
\includegraphics[width=1.\linewidth]{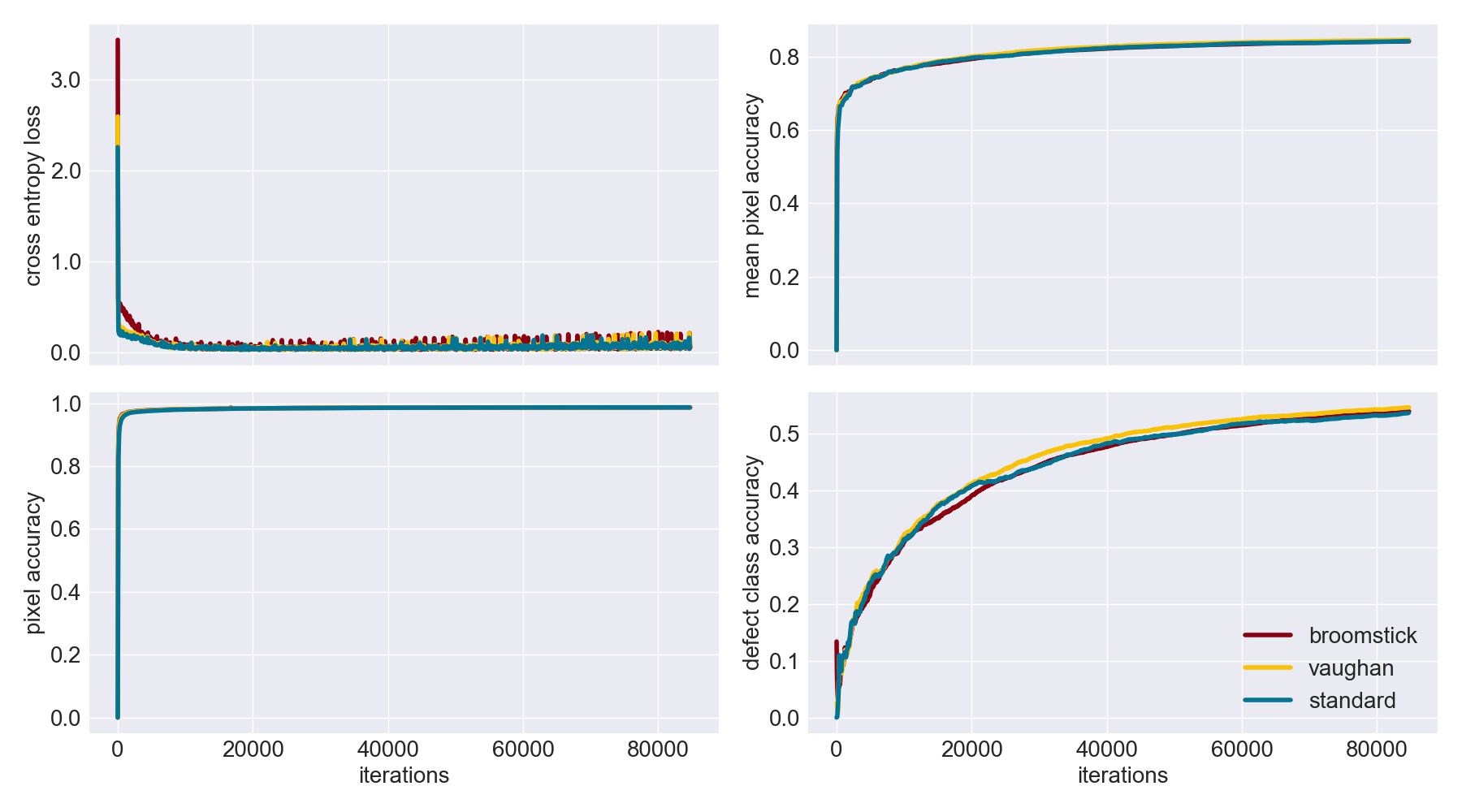}
\caption{Comparison of loss, pixel accuracy, mean pixel accuracy and defect class accuracy of the validation data-set of all three architecture variants. While the overall network accuracy approaches its final value after the first few steps (over 90\,\% PA at step 2,400) defect class accuracy is still rising. Also, defect class accuracy reaches only about 55\,\% accuracy while pixel accuracy approaches 99\,\%. This discrepancy is caused by our highly unbalanced data-set where defects represent only a very small proportion of all pixels.}
\label{fig:Architekturvergleich}
\end{figure*}

\paragraph{Weighted Loss Calculation}
Plotting the metrics (figure \ref{fig:Architekturvergleich}) of our three network architectures reveals the influence of the unbalanced class categories: pixel accuracy, mean pixel accuracy and loss approach their respective maximum (minimum) within the first 10,000 iterations, whereas the defect class accuracy keeps increasing over the whole training, not exceeding 55\,\%. The obvious solution of prolonging network training would result in overfit, as can be observed by the slight but steady rise in validation loss towards the end of training. To nevertheless equalize the influence of all three class categories we introduce weighted defect class loss, additionally to an overall loss weight of 100. 

% visualise_architectures_paper2.py

\begin{figure*}
\centering
\includegraphics[width=1.\linewidth]{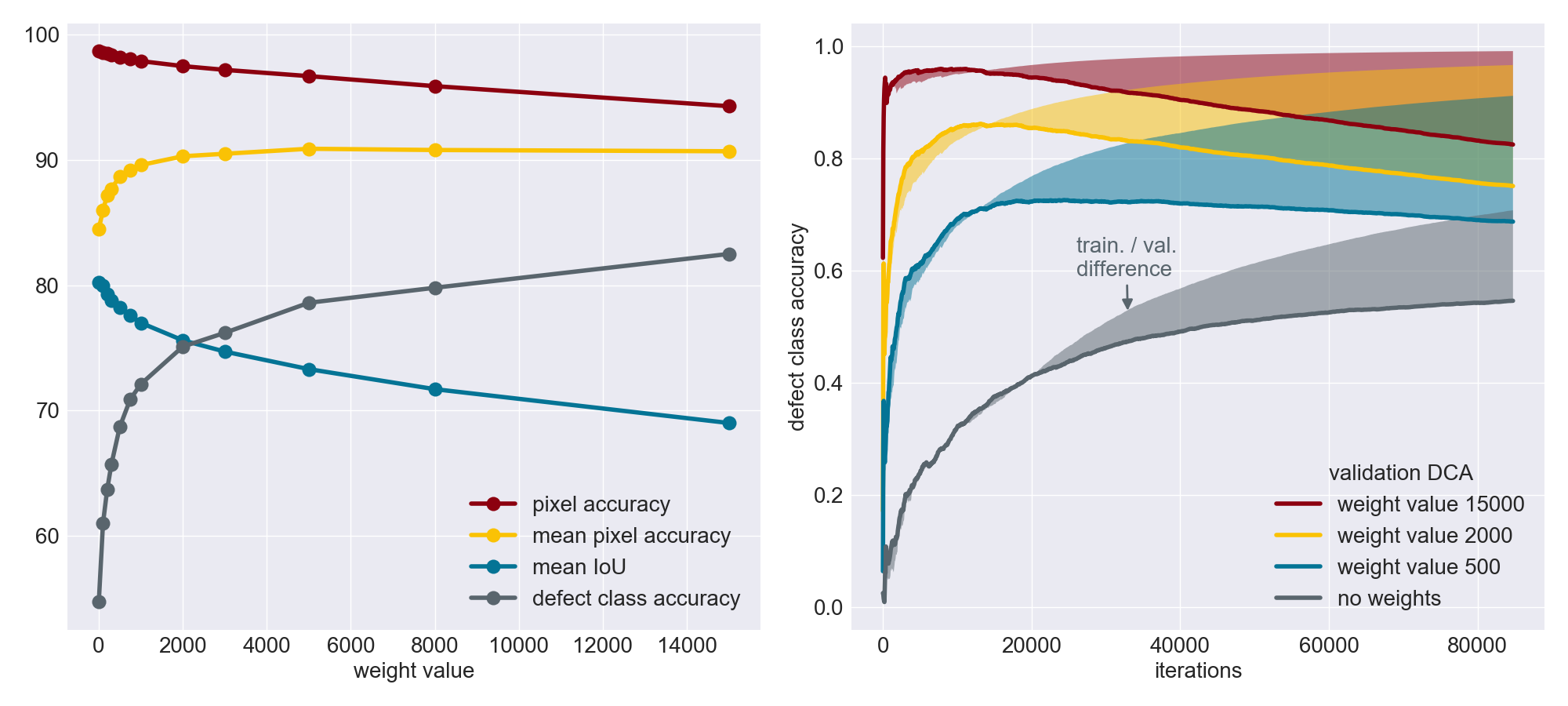}
\caption{Performance metrics of our network trained with varying loss weight values, in order to equalize the unbalance of the data-set. The graph on the left side shows the development of pixel accuracy, mean pixel accuracy, mean IoU and defect class accuracy on the validation data-set over increasing weight values. It becomes apparent that up to a weight value of 2,000 defect class accuracy rises sharply along with mean pixel accuracy. Overall pixel accuracy and mean IoU decrease with increasing weight values as the number of misclassifications in the other two classes rises. The graph on the right side visualizes the development of defect class accuracy during training with regard to different weight values. The lines represent validation defect class accuracy while the area over the lines indicates the deviation from training accuracy. The results show that increasing weight values do correspond with higher defect class accuracy but also give rise to overfit---the network starts to memorise training data and thereby loses its ability to generalise to previously unseen validation data, resulting in increasing training and decreasing validation accuracy.}
\label{fig:weightedNetworks}
\end{figure*}

\begin{figure}
\centering
\includegraphics[width=1.\linewidth]{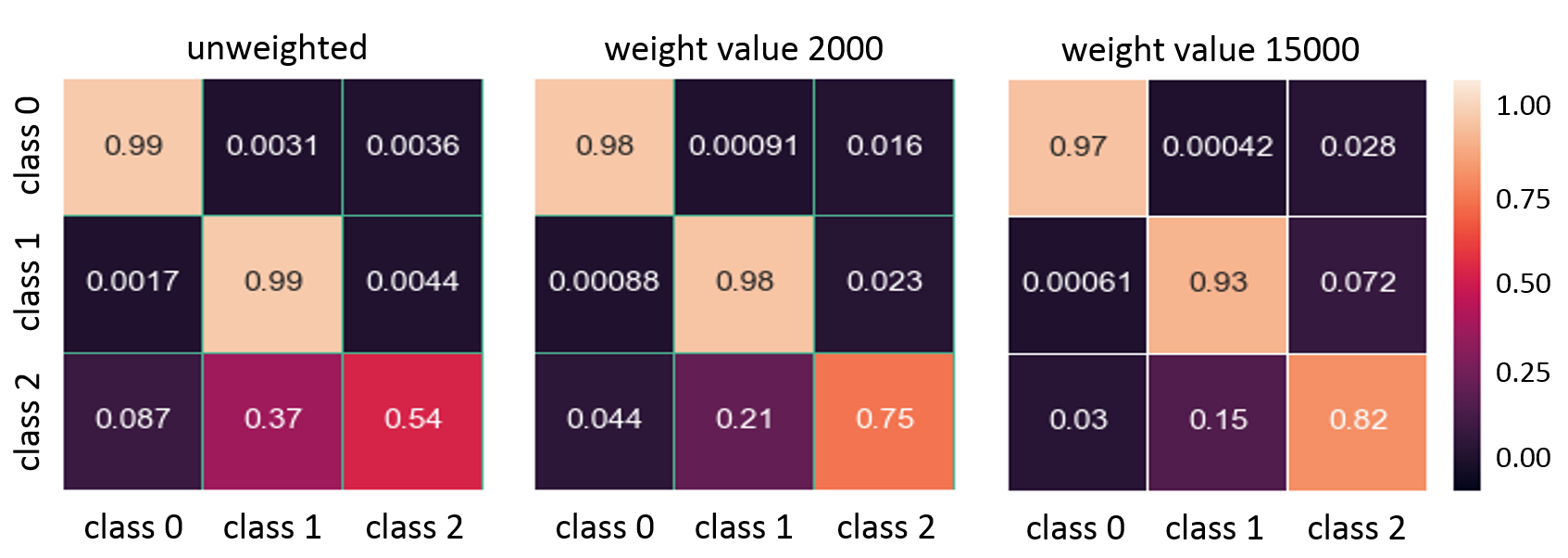}
\caption{Confusion matrices of our network trained with different loss weight values. The first matrix shows an equal weight of 100 for all classes. Here, in-spec chips (class 1) and background pixels (class 0) achieve 99\,\% true positives while defect class reaches only 54\,\% true positives. With increasing loss weight values defect class true positives rise along with an increase of class 1 and class 0 false negatives in favor of defect class.}
\label{fig:konfusionsmatrizen}
\end{figure}

Figure \ref{fig:weightedNetworks} (left) illustrates the development of pixel accuracy, mean pixel accuracy, mean IoU and defect class accuracy, when applying increasing weights on the defect class loss calculation. The correlation of pixel accuracy decrease and defect class accuracy increase (see Table \ref{tab:weightedLoss}) suggests that the network's rate of pixel misclassification in favor of the defect class rises, as is supported by the confusion matrices in figure \ref{fig:konfusionsmatrizen}. While we may ignore misclassifications in class 0 (background) as they do not correlate to actual LED chips, false negatives with regard to class 2 (in-spec chips) might lead to incorrectly rejected LED chips. Depending on the application, false negatives with regard to class 3 (defects) might cause the subsequent rejection of sophisticated packages and thereby high expenses. The selection of a suitable weight value thus ought to include a financial evaluation of misclassification costs. In a mere metrics based selection, we chose a weight value of 2000, where mean pixel accuracy has approached its maximum with a value of 90.3\,\%, pixel accuracy drops to 97.5\,\% and mean IoU to 75.6, while defect class accuracy reaches 75.1\,\%. As illustrated in figure \ref{fig:weightedNetworks} (right), weight values past 500 result progressively in overfitting as training accuracy rises while validation accuracy decreases, making an additional case for moderate weight values. 

% visualise_weightedNetworks_paper3.py

\begin{table}
\centering
% table caption is above the table
\caption{Comparison of the network's performance with different loss weights utilized in order to equalize our heavily unbalanced data-set. We apply a loss weight of 100 to all classes as well as an additional weight of 500, 2,000 and 15,000, respectively, to the defect class. As expected, defect class accuracy rises with increasing weight value, while pixel accuracy decreases slightly. Mean pixel accuracy, on the other hand, improves from 80.2\,\% to about 90\,\%, where it settles down at weight value 2,000 (see also figure \ref{fig:weightedNetworks}). }
\label{tab:weightedLoss}       % Give a unique label
% For LaTeX tables use
\begin{tabular}{lllll}
\hline\noalign{\smallskip}
network & PA  & mIoU & MPA & DCA \\
\noalign{\smallskip}\hline\noalign{\smallskip}
-  & 98.7 & 80.2 & 84.5 & 54.7 \\
500  & 98.2 & 78.2 & 88.7 & 68.7 \\
2,000 & 97.5  &  75.6 & 90.3  & 75.1  \\
15,000 & 94.3  & 69.0  & 90.7  & 82.5 \\
\noalign{\smallskip}\hline
\end{tabular}
\end{table}

\begin{figure*}
\centering
\includegraphics[width=1.\linewidth]{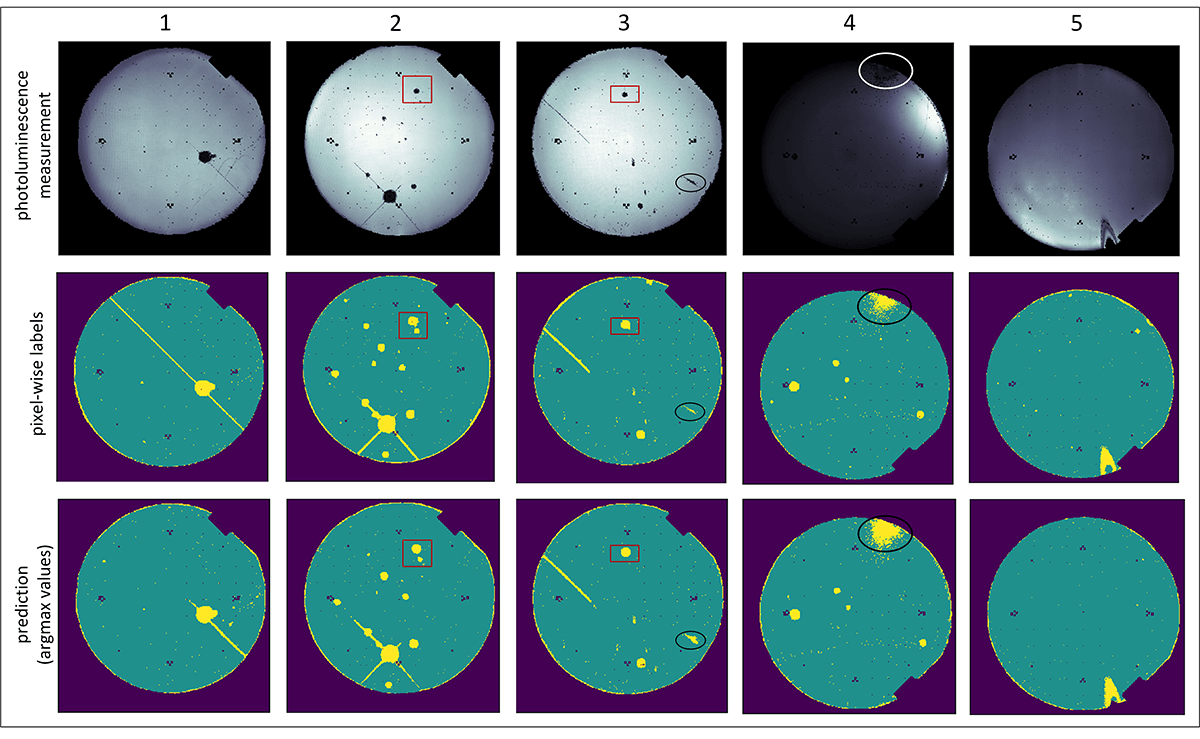}
\caption{Photoluminescence images, labels and prediction results. Columns 1 to 4: Uncommon discrepancies between labels and images are ignored by the network (column 1), while common discrepancies are adapted (columns 2-3, red squares), a behavior that can be transferred to other defect types (columns 3 and 4, black circles). Column 5: Defect clusters of a familiar shape result in an accurate prediction.}
\label{fig:ergebnisse1}
\end{figure*}

\paragraph{Validation Image Analysis}
As described in section \ref{sec:methods}, metrics alone might not always be fully representative of the network's performance, due to the nature of our label acquisition. Because we subsume all available defect causes generated by wafer probing, the labels also include data evaluations as well as results of measurements prior to probing, such as ultrasonic measurements. Hereby determined structural damages are often but not always visible on photoluminescence images, as shown in figure \ref{fig:ergebnisse1}, column 1: while the labels form a continuous line over the whole wafer, only half the line is visible on the photoluminescence image -- which is classified correctly by the network. The second column depicts the difference between the labeled void area and the corresponding, smaller dark spot on the photoluminescence image (red square). It becomes apparent that the network has learned that visible linear defects usually match the labels (column 2 and 3), while voids usually occupy a larger area than visible (column 2 and 3, red squares). That is, the network adopts recurrent mismatch patterns and ignores others. This behavior is also transferred to other defect types with a similar appearance, as depicted in column 3 (circle). Here, the predicted defect area is enlarged as well, while the labels reproduce the visible damage area on the photoluminescence image. Examining the wafer in column~4 could lead to the assumption that the network's enlarged cluster area prediction is caused by the same effect; after all the cluster shape remotely resembles a void. Clusters with a different shape, on the other hand, are predicted correctly (column 5). 

\begin{figure*}
\centering
\includegraphics[width=1.\linewidth]{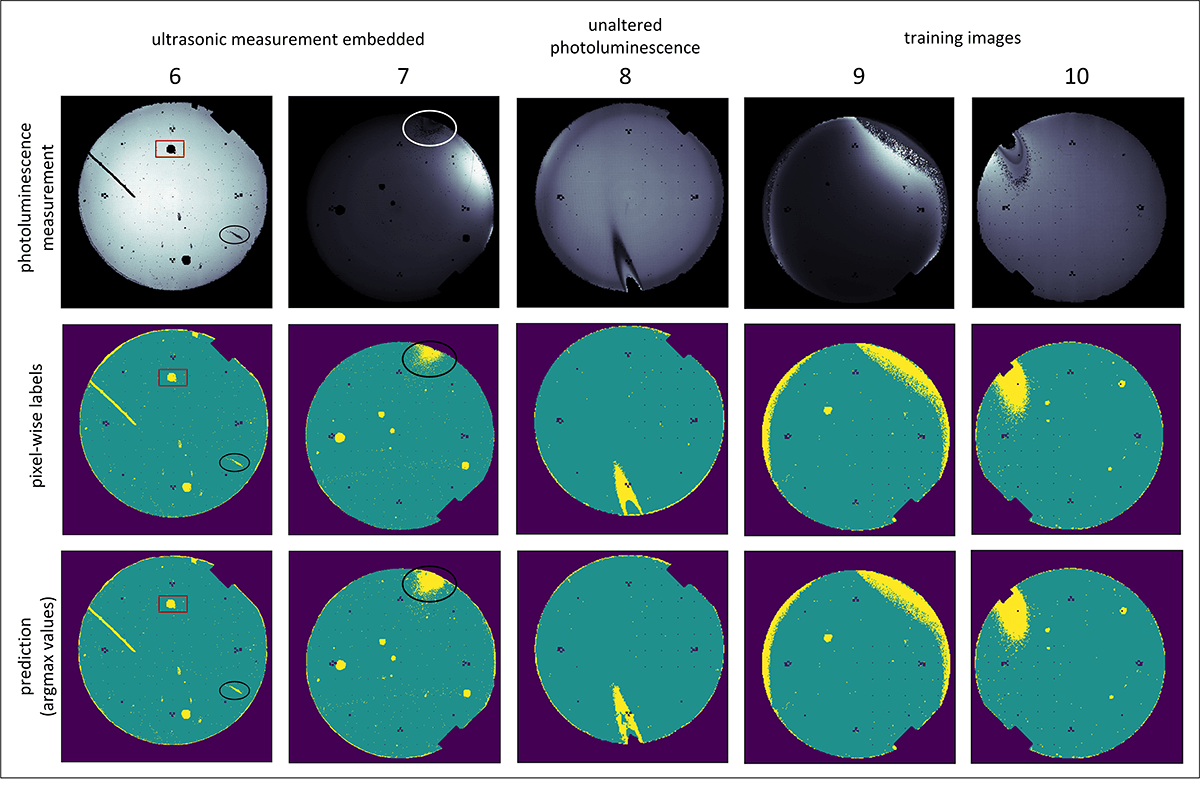}
\caption{Photoluminescence images, labels and prediction results of validation images (6-8) and training images (9, 10). Columns 6 - 8: Embedding photoluminescence images with ultrasonic measurements improves prediction accuracy by removing image-label mismatches (column 6, red square, black circle), although not all misclassifications can be prevented (column 7, black circles), especially those of uncommon defect cluster shapes (column 8). Columns 9 and 10: Evaluating training images reveals that the network is indeed capable of depicting cluster defects accurately, which indicates that the number of training images is insufficient for generalization.}
\label{fig:ergebnisse2}
\end{figure*}

We therefore embedded the photoluminescence images with ultrasonic measurement results and re-trained the network so as to observe possible prediction changes. Columns 6 and 7 (figure \ref{fig:ergebnisse2}) display the same two wafers as columns 3 and 4 (figure \ref{fig:ergebnisse1}), but here training images include ultrasonic measurements. Because now voids occupy the same area on both, photoluminescence and label image, the network no longer enlarges defect areas (figure \ref{fig:ergebnisse2}, column 6, square and circle). Due to the resulting, more precise prediction of ultrasonic defects and a better accordance of photoluminescence image and labels, the overall defect class accuracy rises from 75,1\,\% to 88.7\,\%. However, the defect cluster prediction (column 7, circle) still covers a larger area than photoluminescence image and labels do. We therefore assume that the network has indeed learned to generalize from the defect clusters it has been trained on but struggles with regard to pixel-precise predictions of uncommon or unseen defect cluster shapes. As described before, defect clusters occur rarely in LED manufacturing, resulting in a data-set including only 45 wafers with defect clusters of very different size and shape. As an instance, a defect cluster of such elongated shape as depicted in column 8 is present only once in the data-set (training and validation). While the network reliably detects the defect cluster, the cluster representation itself is holey and shows a fringed border. Our assumption is also supported by the examination of training images (columns 9 and 10). Here, very different defect patterns are predicted in precise accordance with the labels. The degree of accordance suggests, however, that the network has memorized all training cluster shapes, that is it overfits. Extending the data-set with additional cluster wafers would therefore improve prediction accuracy especially with regard to cluster representations. 

\begin{figure}
\centering
\includegraphics[width=1.\linewidth]{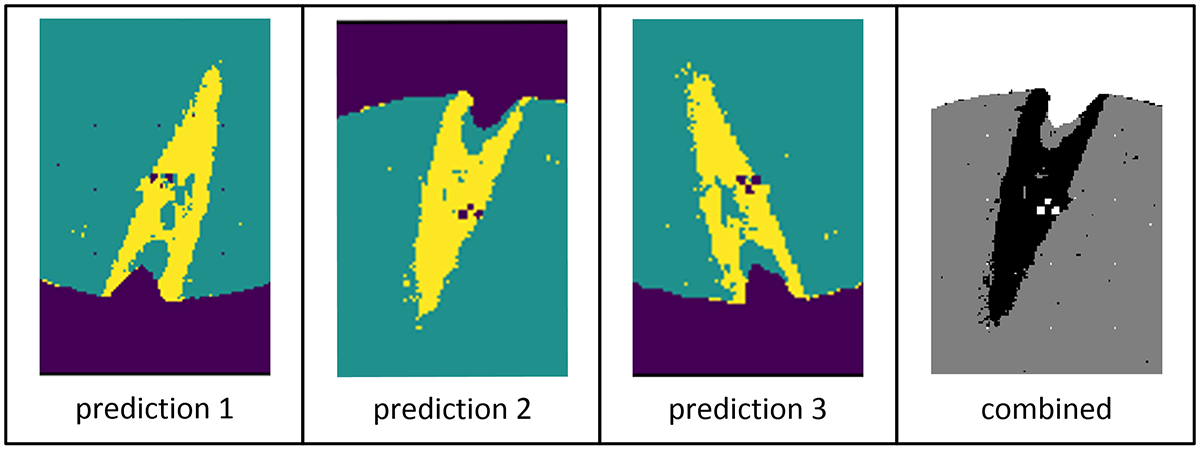}
\caption{Combining the prediction results of differently rotated photoluminescence images can lead to a more precise prediction of cluster defects. The first three images illustrate that the network predicts differently flawed cluster areas for different angles. The combination of all three images is then a more accurate prediction.}
\label{fig:ergebnisse3}
\end{figure}

Another way of improving the pixel-wise prediction of defect clusters (to a certain extend) emerges when analyzing prediction results in both the original orientation as well as rotated (figure \ref{fig:ergebnisse3}). The network handles those three cases differently well. We can use this observation to improve prediction accuracy by running the wafer in varying rotations through the algorithm and combine the results (figure \ref{fig:ergebnisse3}, right image). 

\paragraph{Cross Validation}
In order to confirm the performance of the main experiment we performed an additional \mbox{4-fold} cross-validation. The details and results have been summarized in table \ref{tab:CrossValidation}. With regard to pixel accuracy, mean IoU and mean pixel accuracy, the network achieves similar validation metrics independently from data-set compilation. The prediction accuracy of the defect class varies, however, depending on the data-set. Because the data-set contains only 54 wafers with clusters distinctively deviating in size and shape, we manually split the original data-set so as to equally distribute cluster wafers with regard to quantity as well as diversity. It becomes apparent that less well-balanced data-sets provide not enough examples for the network to generalize from for new defect cluster shapes. Cross-validation fold~3, for example, contains 10 cluster wafers in the validation data-set instead of 8, and as a consequence achieves the lowest defect class accuracy. 

\begin{table}
\centering
% table caption is above the table
\caption{Comparison of validation metrics (pixel accuracy (PA), mean IoU, mean pixel accuracy (MPA) and defect class accuracy (DCA)) based on differently created data-sets. The first data-set ("original") was assembled manually so as to ensure that the training and validation data-set include an equal portion of wafers with defect cluster (46 / 8) as well as well-balanced cluster shapes. The following four data-sets were randomly generated by cross validation. The deviating values in regard to defect class accuracy indicate that hand-picking small, unbalanced data-sets is advantageous for the network's defect classification performance.}
\label{tab:CrossValidation}       % Give a unique label
% For LaTeX tables use
\begin{tabular}{lllll}
\hline\noalign{\smallskip}
data & PA  & mIoU & MPA & DCA \\
\noalign{\smallskip}\hline\noalign{\smallskip}
original  & \textbf{97.5} & 75.6 & \textbf{90.3} & \textbf{75.1} \\
fold 1  & 97.4 & \textbf{76.9} & 89.2 & 72.8 \\
fold 2  & 97.4 & 75.8 & 89.4 & 71.6 \\
fold 3 & 97.2  & 76.7  & 88.5  & 70.9  \\
fold 4  & 97.1 & 75.3 & 88.7 & 73.7 \\
\noalign{\smallskip}\hline
\end{tabular}
\end{table}

\section{Conclusion}
\label{sec:conclusion}
\begin{sloppypar}
We have developed a fully convolutional network for chip-wise defect classification of LED chips, based on photoluminescence measurements. Our experiments show that pixel-wise labels, derived by wafer probing, are suitable for network training even though discrepancies between photoluminescence images and defect labels could be observed. Due to the employment of residual shortcuts and skip connections, the network detects extensive defect clusters as well as single defects. Applying weighted loss allowed us to improve defect class accuracy from 54.7\,\% to 75.1\,\%, with a pixel accuracy of 97.8\,\%, mean pixel accuracy of 90.3\,\% and mean intersection over union of 75.6. Furthermore, image analysis allowed us to detect the biggest deviation between images and labels and by embedding ultrasonic measurement results in photoluminescence images, we were able to further improve defect class accuracy to 88.7\,\% as well as increase network performance (97.9\,\% pixel accuracy, 78.5 mIoU, 95.0\,\% mean pixel accuracy). We expect additional improvements of network performance by a future extension of our data-set with more wafers displaying cluster defects. Finally, our results show that photoluminescence images, analyzed with fully convolutional networks, enable a highly effective quality control in LED manufacturing. 
\end{sloppypar}

\begin{acknowledgements}
We would like to acknowledge support of our work from the German Federal Ministry of Education and Research (BMBF), as part of the joint project \textit{InteGreat}. Moreover, we would like to thank those who share their knowledge in blogs and patiently answer programming questions of strangers. In particular we would like to thank the authors of numpy groupies. 
\end{acknowledgements}

\bibliography{refs}
%\bibliographystyle{apalike}
% BibTeX users please use one of
%\bibliographystyle{apalike}      % basic style, author-year citations
\bibliographystyle{spbasic}      % mathematics and physical sciences
%\bibliographystyle{spphys}       % APS-like style for physics
% name your BibTeX data base

\end{document}